\documentclass[aps,prl,twocolumn,amsmath,amssymb]{revtex4-2}
\usepackage{graphicx}
\usepackage{color}
\usepackage{xcolor}
\usepackage{bm}
\usepackage{ulem}
\usepackage{amsmath}

\begin{document}
\title{Order metrics of jammed solids: Structures, hyperuniformity, and implications for ultra-stable glasses}


\author{Ding Xu$^{1,\dagger}$}		
\author{Qinyi Liao$^{2,\dagger,\ast}$}
\author{Ning Xu$^{1,\ast}$}
\affiliation{$^1$Hefei National Research Center for Physical Sciences at the Microscale, CAS Key Laboratory of Microscale Magnetic Resonance and Department of Physics, University of Science and Technology of China, Hefei 230026, P. R. China. \\
$^2$Wenzhou Institute, University of Chinese Academy of Sciences, Wenzhou 325000, P. R. China, CAS Key Laboratory for Theoretical Physics, Institute of Theoretical Physics, Chinese Academy of Sciences, Beijing 100190, P. R. China. \\
$^\dagger$ These authors contributed equally to this work\\
$^\ast$To whom correspondence should be addressed. E-mail:  qinyi.liao.phy@gmail.com; ningxu@ustc.edu.cn
}

\begin{abstract}
\noindent  {Due to the lack of long-range order, it remains challenging to characterize the structure of disordered solids and understand the nature of the glass transition. Here we propose a new structural order parameter by taking into account multiple rotational symmetries. By studying its statistics for two-dimensional disordered packings of hard particles along the jamming transition line, we observe the evolution from disordered-particle-rich states to ordered-particle-rich states with the increase of packing fraction, together with the unusual non-monotonic change of the degree of hyperuniformity. At the high packing fraction end of the jamming transition line, the packings are mostly composed of ordered particles and are nearly hyperuniform beyond a finite length. Our work links the local order fluctuations to the thermodynamic stability and density fluctuations of disordered solids. Taking advantage of the order parameter, we propose the structural characteristic of ultra-stable glasses: Although globally disordered evaluated by any single symmetry, they are rich of ordered particles and effectively `globally ordered' with crystal-like density fluctuations.}
\end{abstract}

\maketitle

The nature of the glassy states and the glass transition remains one of the most challenging and puzzling problems for the ubiquitous disordered solids \cite{glass_1,glass_2,glass_3,glass_4,glass_5}. The major difficulty arises from the fact that glasses are structurally disordered and conventional order parameters can hardly tell the structrual differences between liquids and glasses. Recently, it has been found that the instantaneous quenching of liquids at different temperatures to local energy minima results in glasses (inherent structures) with different strength of thermodynamic stability \cite{swap_prx,stable_1,stable_2}. In particular, ultra-stable glasses can be obtained from deeply-cooled liquids, which are expected to be the possible candidate to the ideal glass and hence crucial to understanding the long-standing glass transition problem \cite{ultrastable_exp,ultra_stable_1,ultra_stable_2}. It has been shown that ultra-stable glasses possess even qualitatively different mechanical and vibrational properties from normal glasses quenched from high temperatures \cite{brittle-ductile,wlj_dos,oscillitory shear}. There ought to be some structural origins of these differences, which however cannot be identified by conventional order parameters, either.

Hard particles can form glasses as well, but via compression, because temperature is irrelevant to them. Under further compression, hard particle glasses turn to jammed packings via the jamming transition \cite{corey_jamming,nx_pre1,liu_review,torquaro_review}. Relying on the density of the equilibrium liquid states prior to the compression, the jamming transition occurs at different packing fractions. As a result, a jamming transition line ({\textit J}-line) covering a wide range of packing fractions is formed \cite{miyazaki_prl1,LB_scipost,sastry_jline}. Analogous to the temperature-driven glasses, jammed packings in the {\textit J}-line can be seen as inherent structures of hard particle systems obtained from different degrees of cooling. With the increase of packing fraction, the packings should become more stable. The packings at the high packing fraction part of the {\textit J}-line are thus associated with ultra-stable glasses. Jammed packings are typical model systems to understand properties of glasses \cite{xu2007,torquato_glass,wyart_pre}. Revealing the structural characteristics of jammed packings in the {\textit J}-line can thus shed light on the structural characterization of ultra-stable glasses and the understanding of the glass transition.

It has been claimed that jammed packings in the {\textit J}-line follow the same jamming physics \cite{miyazaki_prl1,LB_scipost,sastry_jline}, although their packing fractions differ much. For example, they are all isostatic, i.e., the average coordination number $z=z_{\rm iso}=2d$ with $d$ being the dimension of space, and show the same set of jamming scalings \cite{sastry_jline}. Intuitively, however, some local order should be developed as the compensation for jammed packings of hard spheres to survive at high packing fractions, so there should be some structural evolution along the {\textit J}-line, although the packings remain isostatic. As reported recently \cite{LB_scipost}, with the increase of packing fraction along the {\textit J}-line, the amorphous ordering of three-dimensional (3D) packings represented by the amount of icosahedral structures grows rapidly, together with the monotonic growth of the rattlers and a weak increase of the bond orientational order. There, only two conventionally-concerned frustrated geometries, corresponding to locally and globally densest structures, respectively, are considered. It is not clear yet whether other geometries matter and whether there exists any order metrics to better characterize and understand the competition among different local structures and the resulting phenomena, e.g., density fluctuations and hyperuniformity.

Hyperuniformity is a hot-debated issue of jammed packings closely related to the structure \cite{torquato_pre1,torquato_pr}. Particulate systems with unusually suppressed density fluctuations on large length scales are referred to as hyperuniform, which is signaled by the vanishing density fluctuations in the long wavelength limit \cite{torquato_prl1,LB_compressibility,torquato_prl2}. Clearly, perfect crystals and quasicrystals are hyperuniform \cite{torquato_pr}, but gases and liquids in equilibrium are normally not. Unlike crystals, gases and liquids do not possess long-range structural order, so it seems that hyperuniformity may not favor disorder. However, recent studies have revealed that hyperuniformity could be observed as well in some disordered systems out of equilibrium, e.g., jammed packings of hard spheres \cite{torquato_prl1,torquato_prl2,LB_compressibility,hard_one-dimension}, periodically driven systems \cite{absorb_hu,absorb_hu2}, and active fluids in circular motion \cite{active_hu}. These hyperuniform systems are usually isotropic as well, which can be utilized to devise materials for potential applications \cite{torquato_pr}, despite the limited understanding of the origin of their hyperuniformity.

It has been conjectured that jammed packings of hard spheres are hyperuniform if they are saturated and strictly jammed \cite{torquato_pre1,torquato_prl1}. Accordingly, possibly the most disordered jammed packings of mono-disperse hard spheres, namely maximally random jammed packings \cite{torquato_prl3}, have been found to meet the condition and hence be hyperuniform \cite{torquato_prl1,torquato_prl2}. However, some simulations give contradictory results \cite{LB_pre1,LB_pre2,olsson_hyper,LB_scipost,moore,liu_2018pre}, leaving the validity of the conjecture under debate. As mentioned above, along the {\textit J}-line, the bond orientational order mildly grows when packing fraction increases \cite{LB_scipost,miyazaki_prl1}. It has also been reported that the 3D jammed packings along the {\textit J}-line become less hyperuniform with the increase of packing fraction \cite{LB_scipost}. These results seem to support that more disordered packings tend to be more hyperuniform. However, note that 3D packings have to face the geometrical frustration, which is absent in two dimensions (2D). Is it possible that such a structural difference leads to different scenarios in 2D packings along the {\textit J}-line?

Here we construct a new order parameter $\delta$ to characterize the local structural order of every single particle in disordered solids. It measures the variation of several bond orientational order parameters with distinct rotational symmetries. We focus on the {\textit J}-line in 2D and  find that there is a crossover packing fraction $\phi_{\rm c}$ separating the evolution from disordered-particle-rich states to ordered-particle-rich states. The special role of $\phi_{\rm c}$ is further manifested by robust changes of multiple quantities across it. In particular, the degree of hyperuniformity shows non-monotonic dependence on packing fraction, being worst at $\phi_{\rm c}$. At the highest accessible packing fraction, the density fluctuations are small and almost independent of wavelength, suggesting that the packings there are nearly hyperuniform beyond a finite length, similar to crystals. These packings are particularly interesting. They are rich of ordered particles defined by $\delta$ and thus `globally ordered' in this sense, although they are globally disordered evaluated by any single symmetry. Our work suggests the structural characteristic of ultra-stable glasses and shows that hyperuniformity in jammed packings can arise from multiple structural origins.

\vspace{3mm}
\noindent {\bf Results}

\begin{figure*}
\includegraphics[width=0.45\textwidth]{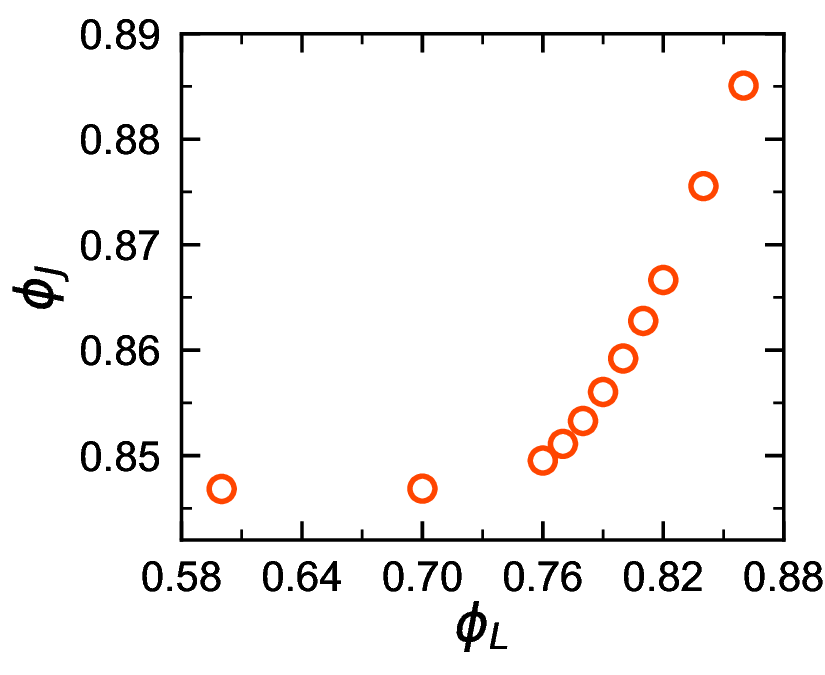}
\caption{\label{fig:fig1} {\bf Formation of the {\textit J}-line from equilibrium liquids}. Correspondence between the packing fraction of jammed packings, $\phi_{\rm J}$, and the packing fraction of equilibrium liquids, $\phi_{\rm L}$, for $N=65536$ systems in 2D. }
\end{figure*}

\noindent {\bf Order parameter.} As described in the Methods section, we obtain the {\textit J}-line by compressing equilibrium hard particle liquids at packing fraction $\phi=\phi_{\rm L}$ to the jamming transition at $\phi=\phi_{\rm J}$. The relation between $\phi_{\rm J}$ and $\phi_{\rm L}$ for 2D systems is shown in Fig.~\ref{fig:fig1}. For particle $j$ in a 2D packing of $N$ particles, we calculate the $l$-fold bond-orientational order (BOO) parameter
\begin{equation}
	\psi_j^l = \frac{1}{n_j}\sum_{k=1}^{n_j}e^{{\rm i}l\theta_{kj}},
\end{equation}
where $n_j$ denotes the number of nearest neighbors of particle $j$ identified by the radical Voronoi tessellation \cite{voro_poly} using Voro++ \cite{voro++}, and $\theta_{kj}$ is the angle between ${\bf r}_{kj}={\bf r}_{k}-{\bf r}_j$ and the $x$-direction with ${\bf r}_k$ and ${\bf r}_k$ being the locations of particles $k$ and $j$. The $l$-fold order of particle $j$ is quantified by $|\psi_j^l|^2$, which is $1$ for perfect $l$-fold bond orientational order and less than $1$ otherwise. In this work, we consider four BOO parameters with $l=5$, $6$, $7$, and $8$. For particle $j$, the four BOO parameters have a mean
\begin{equation}
	\overline{|\psi_j|^2}=\frac{1}{4}\sum_{l=5}^8 |\psi_j^l|^2,
\end{equation}
and a variance
\begin{equation}
	\Delta_j = \frac{1}{4}\sum_{l=5}^8 \left( |\psi_j^l|^2-\overline{|\psi_j|^2} \right)^2.
\end{equation}
Because the four BOO parameters are mutually exclusive, the maximum value of $\Delta_j$ is $\Delta_{\rm max}=0.1875$. We thus define a normalized variance
\begin{equation}
\delta_j=\frac{\Delta_j}{\Delta_{\rm max}},
\end{equation}
which lies in $[0,1]$. For the similar reason, $\overline{|\psi_j|^2}$ should not be a good quantity to characterize the local structural order. For example, a `disordered' particle with all four BOO parameters being close to $1/4$ and $\delta\approx 0$ may have a comparable mean $\overline{|\psi|^2}$ to that of an `ordered' particle with one of the BOO parameters being close to $1$ and $\delta\approx 1$. Therefore, the maximum of the four BOO parameters
\begin{equation}
	O_j ={\rm max}\{ |\psi_j^5|^2,|\psi_j^6|^2, |\psi_j^7|^2, |\psi_j^8|^2\},
\end{equation}
should be more meaningful. Intuitively, a more ordered (disordered) particle should have a larger (smaller) $O$ and a larger (smaller) $\delta$. Then, $O$ and $\delta$ should be correlated, so that both $O$ and $\delta$ can be used as the local order metrics to quantify the order of a particle. We will show that this is exactly the case. By defining these new order parameters, especially $\delta$, we are able to reveal the structural characteristic and refresh our understanding of jammed (disordered) solids associated with their thermodynamic stability.

\vspace{3mm}
\noindent {\bf Evolution of structural order.} In order to form jammed packings of poly-disperse hard particles at higher packing fractions, particles need to pack more efficiently, which involves the competition between local densest packing structures and their spatial organizations. This competition results in complicated spatial fluctuations of local structures \cite{geometry_1,geometry_2,order_1,order_2,order_3,order_4}. Intuitively, there must be some changes of local order and density fluctuations along the {\textit J}-line, even though all the packings maintain isostatic and globally disordered.

\begin{figure*}
\includegraphics[width=0.95\textwidth]{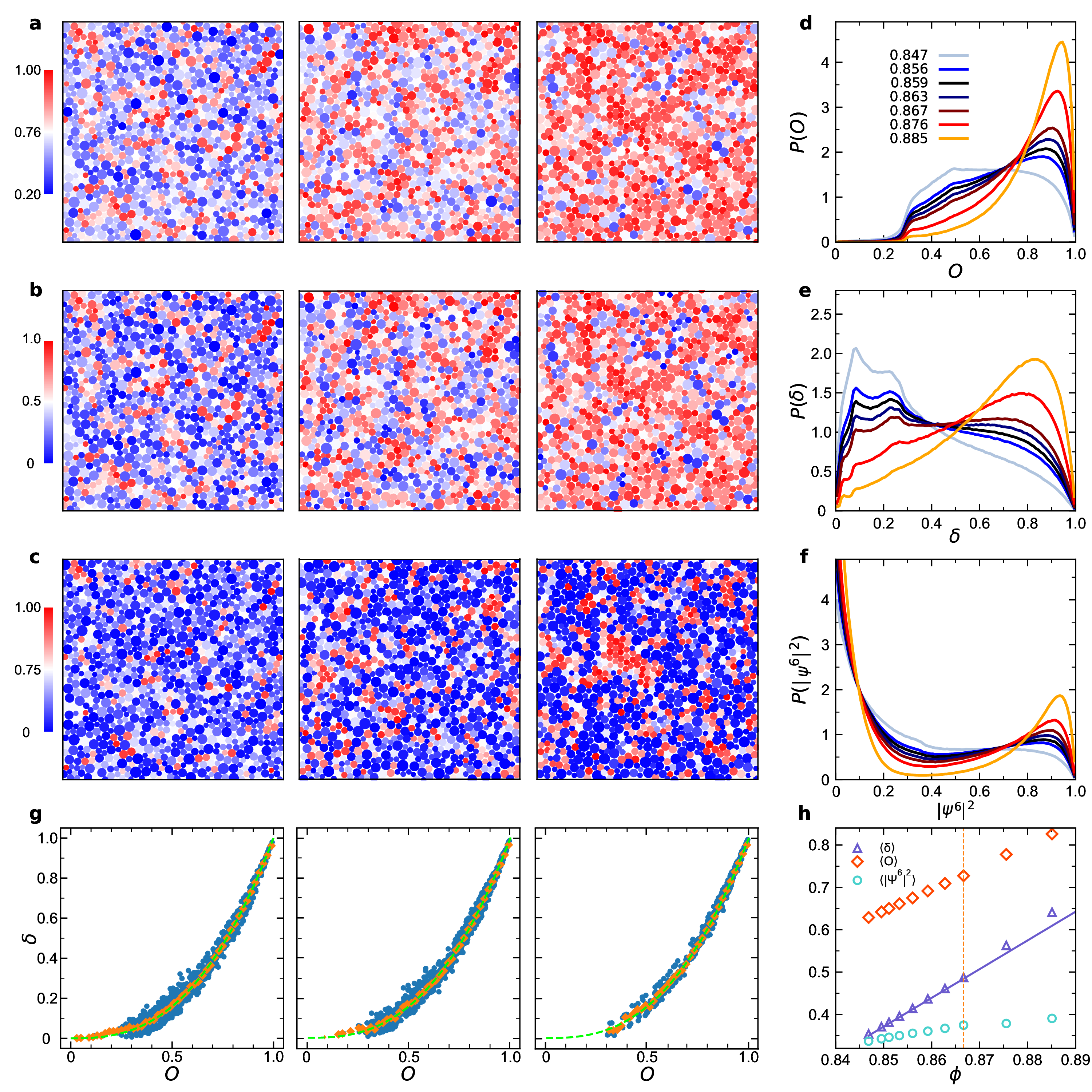}
\caption{\label{fig:fig2} {\bf Evolution of local structure order along the {\textit J}-line.} {\bf a, b, c} Configurations of jammed packings of $N=1024$ particles at $\phi=0.847$, $0.867$, and $0.885$ from left to right, with the color bar showing the maximum BOO parameter $O$, variance of BOO parameters $\delta$, and $6$-fold BOO parameter $|\Psi^6|^2$, respectively. {\bf d, e, f} Evolution of the distributions $P(O)$, $P(\delta)$, and $P(|\Psi^6|^2)$ with packing fraction, respectively, for jammed packings of $N=65536$ particles. {\bf g} Correlation between $O$ and $\delta$ for all particles in configurations shown in {\bf a}, {\bf b}, and {\bf c} (circles). The diamonds are the average over particles. The dashed line shows the relation $\delta=O^{2.65}$. {\bf h} Evolution of $\left< O\right>$, $\left< \delta\right>$, and $\left<|\Psi^6|^2 \right>$ averaged over particles and configurations with packing fraction. The vertical dashed line locates the crossover packing fraction $\phi_{\rm c}$. The solid line is the linear fitting to $\left< \delta\right>(\phi)$ for $\phi<\phi_{\rm c}$. }
\end{figure*}

Figure~\ref{fig:fig2}a-c shows jammed configurations of hard particles at different packing fractions $\phi$ along the {\textit J}-line with every particle being colored according to its maximum BOO parameter $O$, variance of BOO parameters $\delta$, and $6$-fold BOO parameter $|\Psi^6|^2$, respectively. Seen from Fig.~\ref{fig:fig2}a, at the low-$\phi$ end of the {\textit J}-line, most of the particles have low BOO parameters. At the high-$\phi$ end, however, most of the particles are locally ordered. Figure~\ref{fig:fig2}b shows that the spatial distribution of $\delta$ is similar to that of $O$. Figure~\ref{fig:fig2}c indicates that with the increase of $\phi$ the number of particles with a high $|\Psi^6|^2$ also mildly grows, consistent with the 3D results \cite{LB_scipost}, but the direct comparison between Fig.~\ref{fig:fig2}a and \ref{fig:fig2}c indicates that particles with large $|\Psi^6|^2$ only contribute a fraction of ordered particles with large $O$. Although globally disordered viewed from any single BOO parameter, e.g., $\left<|\Psi^6|^2\right>(\phi)$ shown in Fig.~\ref{fig:fig2}h only mildly grows with $\left< .\right>$ denoting the average over particles and configurations, jammed packings at high packing fractions are more like mixtures of locally ordered particles with different symmetries. Let us call such packings as ordered-particle-rich (OPR) states. The low-$\phi$ packings are correspondingly disordered-particle-rich (DPR) states. Therefore, the jammed packings evolve from DPR to OPR with the increase of $\phi$ along the {\textit J}-line.

Figure~\ref{fig:fig2}d-f shows the evolution of the distributions of the three order parameters, $P(O)$, $P(\delta)$, and $P(|\Psi^6|^2)$, with packing fraction, respectively. At the low-$\phi$ end of the {\textit J}-line, $P(O)$ looks flat. With the increase of $\phi$, a peak emerges and grows at high $O$, indicating the growth of the number of ordered particles. Interestingly, the behavior of $P(\delta)$ reminisces the evolution of phase coexistence. At low $\phi$, $P(\delta)$ has a peak and a long tail at low and high values of $\delta$, respectively. With the increase of $\phi$, the peak decays and is eventually replaced with a high-$\delta$ peak at high $\phi$. There is an intermediate $\phi_{\rm c}\approx 0.867$ at which $P(\delta)$ exhibits a plateau. The evolution of $P(\delta)$ more directly illustrates the coexistence of ordered and disordered particles and the evolution from DPR to OPR. The flattening at the intermediate $\phi_{\rm c}$ then signals the crossover between DPR and OPR.  Unlike $P(O)$ and $P(\delta)$, $P(|\Psi^6|^2)$ is bimodal, which is more pronounced with the increase of $\phi$. The growth of the high-$|\Psi^6|^2$ peak indicates the mild growth of ordered particles with the $6$-fold symmetry. Note that the low-$|\Psi^6|^2$ peak contains not only disordered particles, but also ordered particles with symmetries other than $6$, which are all `disordered' evulated only by $|\Psi^6|^2$.

As discussed earlier, there may be some correlation between $O$ and $\delta$. In Fig.~\ref{fig:fig2}g, we show their correlation for all particles of the three configurations in Fig.~\ref{fig:fig2}a-c. In general, $\delta$ increases with $O$ and fluctuates around the relation $\delta=O^{2.65}$. We can thus derive $P(O)=2.65{\delta}^{0.622}P(\delta)$. Therefore, Fig.~\ref{fig:fig2}d can be transformed to Fig.~\ref{fig:fig2}e using this relation.

\begin{figure*}
\includegraphics[width=0.8\textwidth]{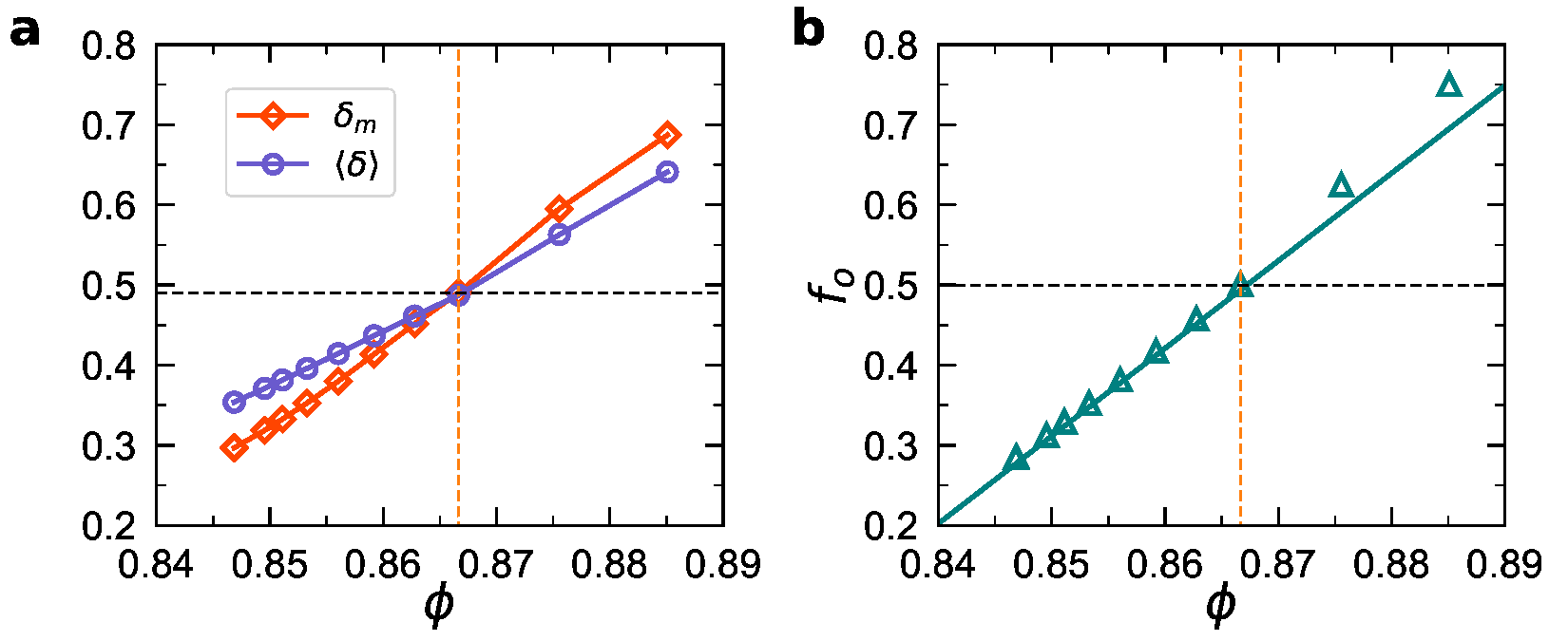}
\caption{\label{fig:fig3} {\bf Quantitative characterization of the evolution from DPR to OPR packings.} {\bf a} Packing fraction dependence of the average and the median of $\delta$, $\left<\delta\right>$ and $\delta_m$ for $N=65536$ systems. The two curves intersect at $\phi_{\rm c}$, from which we determine the crossover value $\delta\approx 0.49$ (horizontal dashed line) to distinguish ordered and disorder particles. {\bf b} Packing fraction dependence of the fraction of ordered particles $f_{\rm o}$. The horizontal dashed line shows $f_{\rm o}=0.5$. The solid line is a linear fit to $f_{\rm o}(\phi)$ at $\phi<\phi_{\rm c}$. In both panels, the vertical dashed line locates $\phi=\phi_{\rm c}$.}
\end{figure*}

Now we see that $O$ and $\delta$ are approximately equivalent to characterize the local order of a particle. An immediate question is where the boundary is to distinguish ordered and disordered particles. The evolution of $P(\delta)$ with $\phi$ shown in Fig.~\ref{fig:fig2}e suggests that the boundary may be defined around $\delta\approx 0.5$. To locate the boundary, we compare the mean and median of $\delta$,  $\left< \delta\right>$ and $\delta_m$, in Fig.~\ref{fig:fig3}a. They agree at $\phi\approx 0.867$ with $\left< \delta\right>=\delta_{\rm c}\approx 0.49$, suggesting that $\delta_{\rm c}$ is a reasonable boundary. Correspondingly, $O_{\rm c}=\delta_{\rm c}^{0.377}\approx 0.76$. Moreover, the crossover packing fraction is also consistent with $\phi_{\rm c}$ associated with the flattening of $P(\delta)$. In previous studies, the choice of the crossover value of the order parameter to determine ordered particles are somehow empirical and casual \cite{order_value_choose_1,order_value_choose_2,order_value_choose_3,order_value_choose_4,order_value_choose_5}. Our $O_{\rm c}$ is close to previously used empirical values \cite{order_value_choose_1,order_value_choose_2}, but our determination of the value is much more definite. This is based upon the definition of the new order parameter $\delta$. We expect that $\delta$ may be a more generalized order parameter to identify local order of particles in various systems and circumstances.

In Fig.~\ref{fig:fig3}b, we show the fraction of ordered particles with $\delta>\delta_{\rm c}$, $f_{\rm o}$, along the \textit{J}-line. It grows with $\phi$ monotonically and reaches $50\%$ around $\phi=\phi_{\rm c}$. Interestingly, $f_{\rm o}(\phi)$ looks linear when $\phi<\phi_{\rm c}$, as demonstrated by the linear fit in Fig.~\ref{fig:fig3}b. When $\phi>\phi_{\rm c}$, $f_{\rm o}(\phi)$ grows faster. We can also see similar changes of $\left< \delta\right>(\phi)$ across $\phi_{\rm c}$ in Fig.~\ref{fig:fig2}h. All these together with more evidence to be shown later indicate that the existence of $\phi_{\rm c}$ is not an accident. At the high-$\phi$ end of the {\textit J}-line accessed in this work, $f_{\rm o}$ and $\left<\delta\right>$ reaches about $75\%$ and $0.83$, respectively. It is difficult to extend the {\textit J}-line to higher packing fractions. However, it can be expected that $f_{\rm o}$ and $\delta$ reach even higher values at higher packing fractions.

Seen from Fig.~\ref{fig:fig1}, the packing fraction of the equilibrium liquid to generate jammed packings at $\phi_{\rm c}\approx 0.867$ is $\phi_{\rm L}\approx 0.820$. Analogous to glasses quenched from different temperatures, this $\phi_{\rm L}$ may correspond to some critical packing fractions of supercooled hard particle liquids, e.g., the mode coupling or the Vogel-Fulcher packing fraction where the relaxation time is expected to diverge \cite{mct,vft_1,vft_2}. Therefore, $\phi_{\rm c}$ separates ultra-stable packings at $\phi>\phi_{\rm c}$ from less-stable packings at $\phi<\phi_{\rm c}$.  Our results thus suggest the structural characteristic of ultra-stable glasses to be examined in various molecular glass-formers. Ultra-stable glasses are OPR states with ordered particles with different symmetries mixing together. Evaluated by any single symmetry, the whole system is globally disordered. However, they are `globally ordered' viewed by the new order parameter $\delta$. To be more ambitious, we expect that the ideal glass, if it exists, is that with the highest possible $\left< \delta\right>$ and $f_{\rm o}$. Next, we will show that this `global order' does result in unusual density fluctuations similar to globally ordered crystals.

\vspace{3mm}
\noindent {\bf Evolution of density fluctuations and hyperuniformity.} Intuitively, ordered particles should have smaller density fluctuations than disordered particles. Along the {\textit J}-line, when more and more particles become ordered, will the density fluctuations be smaller? It is interesting to know if the local order evolution is accompanied with some unusual evolution of density fluctuations.

\begin{figure*}
\includegraphics[width=0.5\textwidth]{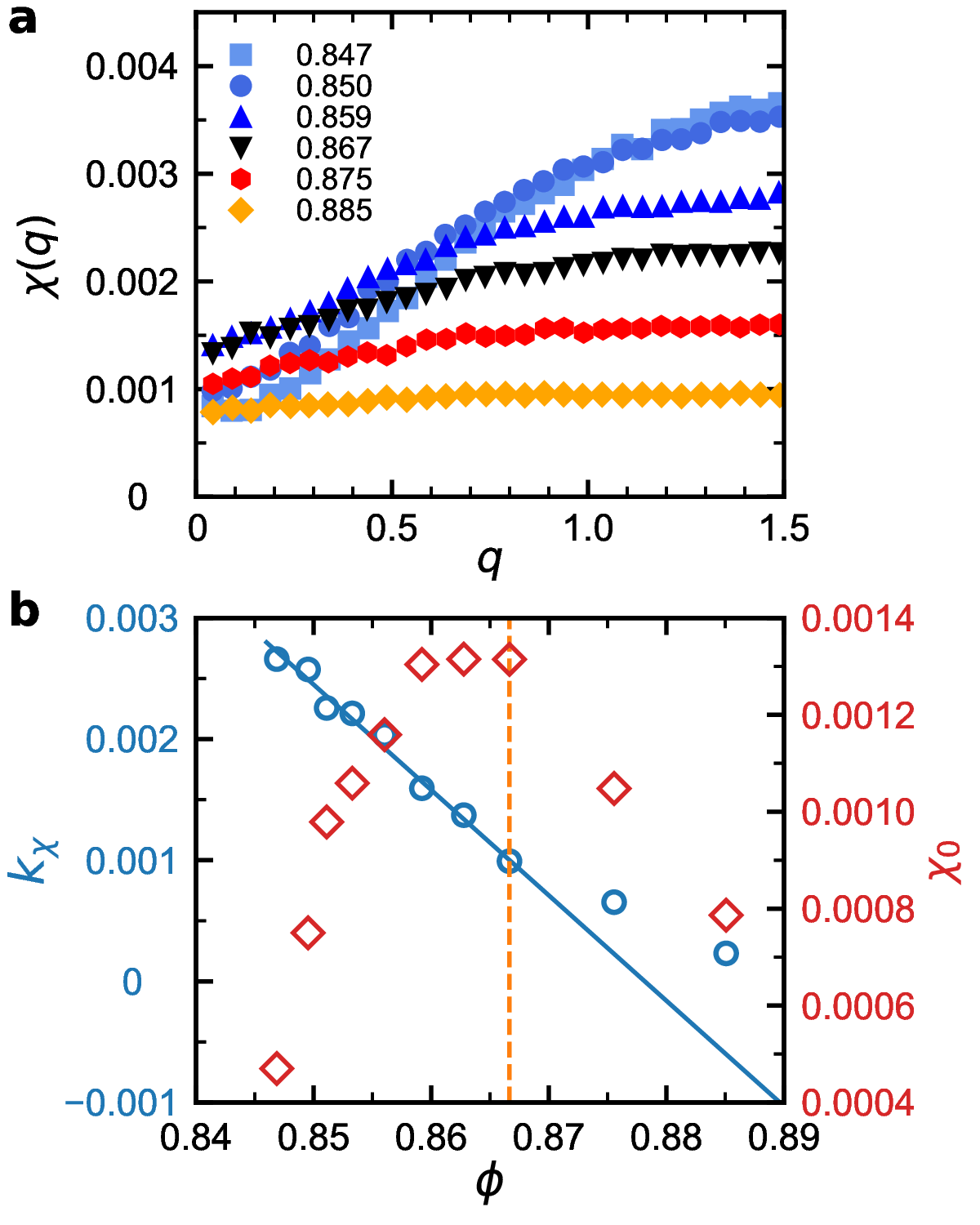}
\caption{\label{fig:fig4} {\bf Evolution of density fluctuations along the {\textit J}-line}. {\bf a} Evolution of the spectral density $\chi(q)$ with packing fraction for $N=65536$ systems. The low-$q$ part can be fitted with $\chi(q)=\chi_0+k_{\rm \chi}q$. {\bf b} Packing fraction dependence of $\chi_0$ (squares) and $k_{\rm \chi}$ (circles), respectively. The vertical dashed line locates $\phi=\phi_{\rm c}$. The solid line is a linear fit to $k_{\rm \chi}(\phi)$ at $\phi<\phi_{\rm c}$. }
\end{figure*}

Figure~\ref{fig:fig4}a compares the spectral density $\chi(q)$ (defined in the Methods section) at different packing fractions along the {\textit J}-line. At the low-$\phi$ end of the {\textit J}-line corresponding to the originally-defined {\textit J}-point \cite{corey_jamming}, the jammed packings are most disordered evaluated by either $\left<|\Psi^6|^2\right>$ or $\left<\delta\right>$. It has been suggested that such packings could be hyperuniform \cite{torquato_prl1,LB_compressibility,torquato_prl2}. The major feature is that $\chi(q)\sim q$ at long wavelengths, so that $\chi(0)\to 0$ and density fluctuations disappear in the $q\to 0$ limit. Like previous studies \cite{LB_scipost,torquato_prl2}, we show that $\chi(q)$ at the low-$\phi$ end of the {\textit J}-line is roughly linear when $q$ is small. However, due to multiple reasons including system size, numerical inaccuracy, etc., it is always difficult to see the true $\chi(0)\to 0$ behavior \cite{torquato_pre2}. Within current precision, we can fit the small-$q$ part of $\chi(q)$ with $\chi(q)=\chi_0+k_{\rm \chi}q$. Smaller $\chi_0$ means better hyperuniformity. $k_{\rm \chi}$ quantifies the wavelength dependence of density fluctuations. Smaller $k_{\rm \chi}$ means that the variation of density fluctuations with the change of length is weaker. For perfect crystals, $\chi_0=0$ and $k_{\rm \chi}=0$ when $q<q_0$ with $q_0>0$ \cite{torquato_pr}.

We fit all $\chi(q)$ curves in Fig.~\ref{fig:fig4}a with the linear relation and show in Fig.~\ref{fig:fig4}b $\chi_0(\phi)$ and $k_{\rm \chi}(\phi)$, respectively. Interestingly, $\chi_0(\phi)$ is non-monotonic. It increases up to a maximum and then decays. The maximum occurs around $\phi_{\rm c}$. With the increase of packing fraction, the packings become less hyperuniform until $\phi\approx \phi_{\rm c}$. Afterwards, they become more and more hyperuniform.  With the assumption that the local densities of ordered and disordered particles are different, this non-monotonic behavior seems to be a natural consequence of the local order evolution. At the high-$\phi$ end of the {\textit J}-line, $\chi_0$ is even comparable to that at the low-$\phi$ end. If the {\textit J}-line can be extended to even higher packing fractions, it is plausible to expect that $\chi_0$ could be even smaller. Therefore, unlike in 3D where $\chi_0$ is shown to monotonically increase and hyperuniformity becomes worse along the {\textit J}-line \cite{LB_scipost}, hyperuniformity may be achieved at the two extremes of the {\textit J}-line in 2D, corresponding to the two limits of low-$\phi$ DPR and high-$\phi$ OPR states, respectively. The lack of geometrical frustration may be the reason for 2D packings to have different $\chi_0(\phi)$ behaviors from 3D ones.

Figure~\ref{fig:fig4}b shows that $k_{\rm \chi}(\phi)$ decreases monotonically. Note that this is also the case for 3D packings \cite{LB_scipost}. Here, $\phi_{\rm c}$ acts as the crossover again. When $\phi<\phi_{\rm c}$, $k_{\rm \chi}(\phi)$ is almost linear. Beyond $\phi_{\rm c}$, $k_{\rm \chi}(\phi)$ apparently departs from the linear behavior. At the high-$\phi$ end of the {\textit J}-line, $k_{\rm \chi}$ decays almost to $0$. Therefore, the packings there are rather particular. They are nearly hyperuniform not just in the $q\to 0$ limit, but beyond some finite length. This behavior is quite similar to crystals. It is surprising that disordered solids can exhibit similar density fluctuations to crystals. However, this counter-intuitive result somehow reflects the effectiveness of our argument that ultra-stable packings are ordered from the perspective of $\delta$.

\vspace{3mm}

\noindent {\bf Discussion}

\noindent  Relying on the new order parameter $\delta$, we find apparent structural evolution from DPR to OPR along the {\textit J}-line and propose the structural characteristic of ultra-stable glasses. They are mixtures of locally-ordered particles with different symmetries and hence `globally-ordered' viewed from $\delta$. This evolution suggests the connection between the thermodynamic stability and the new structural order $\left<\delta\right>$ for disordered solids. More stable solids contain more ordered particles and are more ordered. In 2D, this evolution is also accompanied with the unusual non-monotonic change of the degree of hyperuniformity along the {\textit J}-line. It is interesting that the jammed packings with the highest order is nearly hyperuniform beyond a finite length, similar to crystals.

Note that $\left<\delta\right>=1$ for perfect crystals. In this sense, $\delta$ acts as a generalized order parameter to characterize structures of both crystals and disordered solids. This may also imply that ultra-stable glasses with a high $\left<\delta\right>$ could share some similarities with crystals. The crystal-like density fluctuations at the high-$\phi$ end of the {\textit J}-line should be a direct manifestation. We expect to see more in future studies. Ultra-stable glasses are expected to be close to the ideal glass. The ideal glass, if it exists  , should be most stable with the highest $\left<\delta\right>$. From this perspective, the ideal glass acts as the counterpart of crystals. Recently, a method to construct ideal disk packing in two dimensions has been proposed \cite{eric_ideal}. These ideal packings can be viewed as the prototype of the ideal glass, and we anticipate that it will possess the highest $\left<\delta\right>$.

The liquid-crystal transition is a symmetry-broken process, so could be the formation of the ideal glass, when evaluated by $\left<\delta\right>$. However, multiple rotational symmetries need to be considered simultaneously for the ideal glass formation. If this is the case, our findings outline the structural picture for the glass-transition theories, such as the random first order transition theory \cite{rfot,rfot_2}.

Here we only study jammed packings in 2D. To examine the generality of the findings, temperature-driven transition from liquids to glasses should be studied for various glass-formers. As discussed above, geometrical frustration results in some different behaviors in 3D jammed packings. For example, the hyperuniformity is found to be worse along the {\textit J}-line. However, the density fluctuations still show weaker wavelength dependence with the increase of packing fraction, similar to 2D. We expect that the order parameter $\delta$ works for 3D systems as well and the connection between the thermodynamic stability and the structural order still holds. Only that more symmetries need to be included and the calculation of $\delta$ is more complicated.

\vspace{3mm}

\noindent {\bf Methods}

\noindent We study 2D systems consisting of $N$ hard disks in a square box with side length $L$. Periodic boundary conditions are applied to both directions. To suppress crystallization or fractionation \cite{swap_prx}, we use polydisperse mixtures. The particle diameter $\sigma$ is extracted from a continuous distribution  $ P(\sigma)=A\sigma^{-3} $ with $\sigma_{\rm m} \leq \sigma \leq 2.22\sigma_{\rm m}$, where $A$ is the normalization factor. We set the unit of length to be the average particle diameter $\bar{\sigma}$.

We first thermalize the system via an efficient swap Monte Carlo (MC) algorithm \cite{swap_prx,LB_jline_prl} to obtain well-equilibrated liquid states over a broad range of packing fractions, $\phi_{\rm L}\in (0.60,0.86)$. Then we perform a rapid compression of the liquid states to reach a high value of the reduced pressure $Z=p/(\rho k_{\rm B} T)\sim 10^9$ following the algorithm used in Ref \cite{lqy_prx}. Here $p$ and $T$ are pressure and temperature, $\rho=N/L^2$ is the number density, and $k_{\rm B}$ is the Boltzmann constant, which is set to be $1$. Afterwards, we generate static jammed packings at the jamming threshold following the method used in Ref. \cite{LB_scipost,nx_pre1,sastry_jline}. We obtain the {\textit J}-line with the packing fraction $\phi$ ranging from $0.847$ to $0.885$, as shown in Fig.~\ref{fig:fig1}.

To characterize the density flunctuations and hyperuniformity of jammed packings, we calculate the spectral density,
\begin{equation}
	\chi(q) = \frac{1}{L^2}\langle u({\bf q}) u({\bf -q}) \rangle \label{chiq},
\end{equation}
with
\begin{equation}
	u({\bf q}) = \sum_{j=1}^{N}e^{{{\rm i}}{{\bf q}\cdot {\bf r}_j}}f_j({\bf q}),
\end{equation}
where $\langle\cdot\rangle$ denotes the average over configurations, $q=|{\bf q}|$ with ${\bf q}$ being the wave vector in the reciprocal space, and $f_j({\bf q})$ is the shape factor defined as
\begin{equation}
	f_j({\bf q})=\int_{L^2} d^2r e^{{\rm i}{\bf q}\cdot ({\bf r}-{\bf r}_j)}\Theta(\sigma_j/2 - |{\bf r}-{\bf r}_j| ),
\end{equation}
with ${\bf r}$ and $\sigma_j$ being the vector in real space and diameter of particle $j$. Apparently, $f_j({\bf q})$ is the function of particle diameter $\sigma_j$. For mono-disperse particle systems, all particles have the same shape factor, so that $\chi(q)$ is proportional to the static structure factor $S(q)= \left| \sum_{j=1}^N e^{{\rm i}{\bf q}\cdot {\bf r}_j}\right|^2/N$ purely determined by the particle locations. For poly-disperse systems, $\chi(q)$ is jointly determined by particle locations and diameters.

\vspace{3mm}

\noindent{\bf Acknowledgements}

\noindent We thank Ludovic Berthier, Anthony Chieco, Atsushi Ikeda, Yuliang Jin and Misaki Ozawa for helpful discussions. This work was supported by the National Natural Science Foundation of China Grants No. 12334009 and We also thank the Supercomputing Center of University of Science and Technology of China for the computer time.


\vspace{3mm}

\noindent{\bf Author contributions}

\noindent Q. L. and N. X. designed the project. D. X. and Q. L. performed the simulations. D. X., Q. L. and N. X. analyzed the data and wrote the paper. Correspondence and requests for materials should be addressed to Q. L. (email: qinyi.liao.phy@gmail.com) or N. X. (email: ningxu@ustc.edu.cn)

\vspace{3mm}

\noindent {\bf Competing interests}

\noindent The authors declare no competing financial interests.

\vspace{3mm}

\noindent{\bf Data availability}

\noindent The supporting data are available from the corresponding authors upon request.

\end{document}